\documentclass[a4paper,11pt]{article}
\pdfoutput=1 

\usepackage{jcappub} 

\usepackage[T1]{fontenc} 
\usepackage{amssymb}

\title{\boldmath Fermionic Dark Matter in a simple $t$-channel model}

\author{Ashok Goyal\footnote{Visiting researcher}}
\author{and Mukesh Kumar}
\affiliation{National Institute for Theoretical Physics,\\
School of Physics and Mandelstam Institute for Theoretical Physics,\\
University of the Witwatersrand, Johannesburg,\\
Wits 2050, South Africa.}

\emailAdd{agoyal45@yahoo.com}
\emailAdd{mukesh.kumar@cern.ch}

\abstract{We consider a fermionic dark matter (DM) particle in renormalizable Standard Model (SM) gauge
interactions in a simple $t$-channel model. The DM particle interactions with SM fermions is through the 
exchange of scalar and vector mediators which carry colour or lepton number. In the case of coloured 
mediators considered in this study, we find that if the DM is thermally produced
and accounts for the observed relic density almost the entire parameter space is ruled out by the direct
detection observations. The bounds from the monojet plus missing energy searches at the Large Hadron
Collider are less stringent in this case. In contrast for the case of Majorana DM, we obtain strong bounds from the
monojet searches which rule out DM particles of mass less than about a few hundred GeV for both the scalar 
and vector mediators. }

\begin{document}
\maketitle
\flushbottom

\section{Introduction}
\label{sec:intro}

Several astrophysical and cosmological observations like the rotation profile of galaxies, cosmic microwave
background (CMB), large scale structure and type Ia supernovae point towards the existence of Dark Matter
(DM) in the universe. A precise measurement of CMB by the Planck Satellite mission~\cite{Ade:2015xua} 
has determined 
the amount of cold dark matter (CDM) in the universe to be $\Omega_{\rm DM} h^2 = 0.1188 \pm 0.0010$.

Dark matter particle search has emerged as one of the most engaging fields of research in Astro-particle 
Physics. Weakly interactive massive particle (WIMP) dark matter searches look for missing energy signature
of such particles at hadron colliders. Direct detection consists in studying the nuclear - recoil energy and its
spectrum in non-relativistic elastic collision of WIMPS with the atomic nucleus. The indirect detection experiments
need not rely on the presence of DM particles in the vicinity of earth. These searches aim at detecting the signature
of annihilating or decaying DM in cosmic rays. The DM particles $\chi$ will generally annihilate into Standard Model
(SM) particles. The primary decay products will decay further into electrons, positrons, protons, antiprotons, neutrinos
and $\gamma$-rays which can be observed by dedicated detectors.

Many theories beyond the SM predict particles which are neutral, stable and are massive. These are viable DM
candidates notably, neutralino in supersymmetric theories~\cite{Goldberg:1983nd, Ellis:1983ew, Kane:1993td}, 
the lightest Kaluza-Klein particle in theories with
extra dimensions~\cite{Kolb:1983fm, Cheng:2002ej}
or the lightest T-odd (heavy photon) particle in Little Higgs Model~\cite{Birkedal:2006fz}. 
In addition there are effective
field theories (EFT)~\cite{Beltran:2008xg, Yu:2011by, Goodman:2010yf, Goodman:2010ku, Goodman:2010qn, Fox:2011pm} 
in which the DM-SM interactions are mediated by some superheavy mediator particles which
exist at some new high energy scale $\Lambda$. Such theories have been analysed in detail using the Large
Hadron Collider (LHC) Run I data. There exists another class of theories called simplified DM models in which
the state mediating between the DM and SM particles plays an important role. Such theories in contrast to the 
EFT's are able to decide the full kinematics of DM production at the LHC. Recently the desirability of building such
simplified models and the criteria they should satisfy has been discussed in detail by the authors of 
Ref.~\cite{Abdallah:2014hon}.

In this paper we consider a minimal vector-like baryonic/leptonic spin-1/2 DM with renormalizable SM gauge
interactions in a simple beyond the SM scenario. The DM particle $\chi$ in this case is a $t$-channel 
annihilator. Its interaction with the SM particles is through the exchange of lepto-quark type spin-0 (S) or a
spin-1 vector (V) particle. A class of such models for scalar mediator coupling has been considered in the
literature~\cite{Chang:2013oia, DiFranzo:2013vra, An:2013xka, Papucci:2014iwa}. 
The mediators S or V carry colour or leptonic index. In section~\ref{model} we 
describe the model. In section~\ref{contraint} we discuss all relevant experimental constraints. 
The relic density contributed by these particles is calculated and 
constraints on the parameters of the model assuming that the contribution of $\chi$'s alone does not
exceed the observed relic density are obtained in~\ref{relic}. 
With these constraints in place, we discuss the compatibility of these
constraints from the direct and indirect detection experiments in~\ref{dir} and~\ref{indir}
respectively. In~\ref{collider} we examine the signature of these DM particles at the LHC where a 
monojet signal with missing energy is investigated. Section~\ref{summ} is devoted to the summary and
discussion of our main results.                  

\section{The Model}
\label{model}
The model consists of a single Dirac vector-like fermion $\chi$ interacting through the mediation of a scalar ($S$)
or a vector boson ($V^\mu$) which carry a baryonic (colour) or leptonic index. The quantum numbers of the DM particle
$\chi$ and mediators along with their transformation properties are given in Table~\ref{tab:i}.

\begin{table}[tbp]
\centering 
\begin{tabular}{|l |c c c c c |l|} \hline
\hline
Particle	& Spin   & SU(3)$_{C}$ &  SU(2)$_{L}$  &  U(1)$_{Y}$  &  ${\mathbb Z}_2$ & Coupling \\ \hline\hline
Dark Matter, $\chi$ & 1/2 & 1  & 1 & 0 & -1 &  \\ \hline
Baryonic Scalar      & 0   & 3  &  2 & 1/6  & -1 & $\bar q_{_L} \chi_{_R} S_{_b}$  \\ 
mediator, $S_b$      & 0   & 3  & 1  & 2/3, -1/3  & -1 & $\bar u_{_R} \chi_{_L} S_{_b}$, $\bar d_{_R} \chi_{_L} S_{_b}$  \\\hline
Leptonic Scalar      & 0   & 1  &  2 & -1/2  & -1 & $\bar l_{_L} \chi_{_R} S_{_l}$  \\ 
mediator, $S_l$      & 0   & 1  & 1  & -1  & -1 & $\bar e_{_R} \chi_{_L} S_{_l}$  \\\hline
Baryonic Vector               & 1   & 3  & 2 & 1/6  & -1 & $\bar q_{_L} \gamma_\mu \chi_{_L} V_b^\mu$  \\ 
mediator, $V_b^\mu$      & 1   & 3  & 1  & 2/3, -1/3  & -1 & $\bar u_{_R} \gamma_\mu \chi_{_R} V_b^\mu$, $\bar d_{_R} \gamma_\mu \chi_{_R} V_b^\mu$  \\\hline
Leptonic Vector               & 1   & 1  &  2 & -1/2  & -1 & $\bar l_{_L} \gamma_\mu \chi_{_L} V_{_l}^\mu$  \\ 
mediator, $V_l^\mu$       & 1   & 1  & 1  & -1  & -1 & $\bar e_{_R} \gamma_\mu \chi_{_R} V_{_l}^\mu$  \\\hline
\hline
\end{tabular}
\caption{\label{tab:i} Particle content with corresponding quantum numbers and interactions in a $t$-channel model by
considering fermionic dark matter candidate $\chi$ interacting with the SM fermions through scalar and vector mediators.}
\end{table}
In general we will have three mediator SU(2)$_L$ doublets or singlets corresponding to three generations of
hadrons and leptons respectively. We have invoked a discrete symmetry $\mathbb Z_2$ under which the new
particles in the model are odd to ensure stability of the DM $\chi$. The interaction Lagrangian for the leptonic
and baryonic DM for scalar and vector can be written as
\begin{align}
{\cal L}_{\rm scalar} = {\cal L}_{\rm scalar}^{\rm baryonic} + {\cal L}_{\rm scalar}^{\rm leptonic}  
  \supset - \sum\limits_{i} c^{b_i}_s \bar {q}_{_L}^i \chi_{_R} S_b^i  - \sum\limits_{i} c^{l_i}_s {\bar l}_{_L}^i \chi_{_R} S_{_l}^i 
  + {\rm h.c.} \label{eqs}
\end{align}
and 
\begin{align}
{\cal L}_{\rm vector} = {\cal L}_{\rm vector}^{\rm baryonic} + {\cal L}_{\rm vector}^{\rm leptonic}  
  \supset - \sum\limits_{i} c^{b_i}_v \bar {q}_{_L}^i \gamma_\mu \chi_{_L} V_b^{\mu i}  - \sum\limits_{i} c^{l_i}_v \bar {l}_{_L}^i \gamma_\mu \chi_{_L} V_{_l}^{\mu i} + {\rm h.c.} \label{eqv}
\end{align}
where $i$ runs over the three generations and the colour index has been suppressed. 
We have similar interactions for SU(2)$_L$ singlet mediators which couple to right-handed quarks and leptons. The
leptonic and hadronic cases can be treated separately. 

The scalar and vector mediators carry SM charges and would therefore interact with SM gauge bosons.
The relevant interaction Lagrangian for the baryonic can be written as
\begin{equation*}
{\cal L}_{G} = {\cal L}_{G, scalar}^{baryonic} + {\cal L}_{G, vector}^{baryonic}  
\end{equation*}  
where $G$ stands for all SM gauge bosons and
\begin{align}  
{\cal L}_{G, scalar}^{baryonic} =&\, \left(D_\mu S_b \right)^\dag \left(D^\mu S_b \right) - m_S^2 S_b^\dag S_b, \label{lgs} \\
{\cal L}_{G, vector}^{baryonic} =&\, -\frac{1}{4} \left( V_b \right)_{\mu\nu}^\dag \left( V_b\right)^{\mu\nu}
+ m_V^2 \left( V_b\right)_\mu^\dag \left( V_b\right)^\mu
+ i g_s \left(V_b\right)_\mu^\dag t^a \left( V_b\right)_\nu G_a^{\mu\nu}. \label{lgv}
\end{align} 
 The covariant derivative is 
 $D_\mu = \partial_\mu + i g_s t_a G^a_\mu + i g \tfrac{1}{2}{\vec{\tau}}\cdot{\overrightarrow{W}}_\mu + i g^\prime \tfrac{1}{2} Y B_\mu$, 
 $\left(V_b\right)_{\mu\nu} = D_\mu \left(V_b\right)_\nu - D_\nu \left(V_b\right)_\mu$ and $g_s$ is the QCD strong
 coupling.  
 The decay width of scalar and vector baryonic mediators $\Gamma\left( S_b^i/V_b^i \to \chi \bar f_i \right)$ (where $i$ 
 is the generation index) is give by
 \begin{align}
 \Gamma \left(S_b^i \to \chi \bar f_i \right) &= \frac{\left(c_s^i\right)^2}{16 \pi m_{S_i}^3} \left(m_{S_i}^2 - m_{f_i}^2 - m_\chi^2\right) 
 \nonumber 
 \\&\qquad\qquad\qquad \times 
 \left\{m_{S_i}^4 + m_{f_i}^4 + m_\chi^4 - 2 m_{S_i}^2 m_{f_i}^2 - 2 m_{S_i}^2 m_\chi^2 - 2 m_\chi^2 m_{f_i}^2 \right\}^{1/2} \nonumber \\
 &\simeq \frac{\left(c_s^i\right)^2}{16 \pi} m_{S_i} \left( 1- \frac{m_\chi^2}{m_{S_i}^2}\right),
 \end{align}
 since $m_{S_i}, m_\chi \gg m_{f_i}$ is true for all quarks except top-quark. The corresponding decay width of the
 vector mediator is given by
\begin{align}
 \Gamma \left(V_b^i \to \chi \bar f_i \right) &= \frac{\left(c_v^i\right)^2}{24 \pi m_{V_i}^3} \left(m_{V_i}^2 - \frac{m_{f_i}^2}{2}  
 - \frac{m_\chi^2}{2} - \frac{1}{2}\left( m_\chi - m_{f_i}\right)^2 \right) \nonumber \\
 &\qquad\qquad\qquad \times 
 \left\{m_{V_i}^4 + m_{f_i}^4 + m_\chi^4 - 2 m_{V_i}^2 m_{f_i}^2 - 2 m_{V_i}^2 m_\chi^2 - 2 m_\chi^2 m_{f_i}^2 \right\}^{1/2} \nonumber \\
 &\, \simeq \frac{\left(c_v^i\right)^2}{24 \pi} m_{V_i} \left( 1- \frac{m_\chi^2}{m_{V_i}^2}\right)^{3/2}.
 \end{align}
In the above expressions $m_{S_i}$ and $m_{V_i}$ are the masses of the baryonic scalar and vector mediators respectively.
We have similar expression for the decay width of the leptonic mediator where $i$ runs over the three generations of leptons.
The leptonic and baryonic case can be treated separately. Here we will concentrate on the baryonic case only.
Majorana DM with scalar lepton mediator has been considered in Ref.~\cite{Baek:2015fma}.

Unlike the $s$-channel mediator where a single vector boson is required as a mediator, in a $t$-channel model
one requires a different mediator for left-handed quark doublets and right-handed quark singlets of each
generation. In general the interactions given in Lagrangians~\ref{eqs} and~\ref{eqv}  
induces flavour-changing neutral currents which are strongly constrained by low energy phenomenology. However, the constraints can be avoided by imposing minimal flavour violation (MFV) structure on the Yukawa couplings. Flavour violation can also be avoided by restricting to one generation and taking the masses of mediators in the SU(2)$_L$ doublet to be degenerate. Lepton flavour violation can likewise be circumvented. Thus for MFV, we consider the universal coupling with all generations and take the masses of all three generations of mediators to be equal. The case of SU(2) singlet mediators is similar to the SU(2)$_L$ doublet case except that there are more parameters. In what follows we will consider the SU(2)$_L$ doublet case for the sake of simplicity. We thus have only two parameters - the universal coupling $c_s$ ($c_v$) and the common mediator mass $m_S$ ($m_V$) for scalar (vector) mediators where the generation index $i$ and the subscripts $b/l$ from the couplings and massed have been dropped. 

The simplified $t$-channel vector mediator model considered in this study should ideally be embedded into a 
UV-complete theory. The $t$-channel vector and scalar mediators carry flavour index unlike the $s$-channel
mediators where a single vector or scalar mediator is required. The UV-complete theory in the $s$-channel case can be
constructed by extending the SM gauge group to include a new U(1)$^\prime$ symmetry which is spontaneously
broken such that the mediator obtains a mass. In the $t$-channel vector mediated model, flavour symmetry has to
be gauged and unified with SM gauge groups.
The $t$-channel vectors must get their mass from the spontaneous breaking of a unified group down to the SM.
And the $t$-channel mediators of interest would then correspond to the broken, off-diagonal generators of that unified
group. 
However it may be difficult to achieve particularly if MFV is to be
incorporated. Alternatively a UV-complete theory with scalar or vector mediators may be relatively easier to 
implement by endowing the flavour quantum number to the DM particles thereby releasing the mediators from the
burden of carrying flavour and the need for gauging the flavour symmetry. In this case the mediators may
carry colour as well as weak charges depending on whether it couples to DM and quarks or DM and leptons. 
The flavoured DM models with the implementation of MFV have been considered in
the literature (see for example~\cite{Kile:2013ola}). 
The case of coloured scalar mediator $t$-channel model is much like the
minimal supersymmetric model type, where the coloured scalar quark has the same flavour index as the SM 
quark~\cite{Agrawal:2014una}.  
One of the avowed purpose of simplified model approach is to characterise the DM production present
in the UV-complete model without having to specify the entire UV completion. Simplified models contain both the
DM and mediator particles, the latter being the link between the SM and DM. These models obviously do not contain
all the ingredients present in the UV-complete models of dark matter. 
Recently~\cite{Englert:2016joy} it has been shown that at 
arbitrary high energies, the use of simplified models at the LHC and at future colliders are generally valid, in particular
the models which have SM gauge invariance.     
     
\section{Constraints}      
\label{contraint}
In this section we study the experimental constraints on the model parameters $m_\chi, m_S$ and $m_V$ from relic
density, direct and indirect observations and from the collider signals. 

\subsection{Relic density}
\label{relic}  
In the early universe the dark matter particles $\chi$ were light in thermal equilibrium with the rest of the plasma through
the pair creation and annihilation of $\chi$'s. The annihilation process $\chi\bar \chi \to f\bar f$ proceeds through the 
$t$-channel mediators scalar or vector. The thermal relic density of $\chi$'s is obtained by solving the Boltzmann 
equation:
\begin{equation}
\frac{d\eta_\chi}{dt} + 3 H \eta_\chi = - \langle \sigma | v |\rangle \left ( \eta_\chi^2 - {\eta_\chi^{\rm EQ}}^2\right),
\end{equation} 
where 
\begin{equation}
H = \frac{\dot R}{R} = \sqrt{\frac{8\pi\rho}{3M_{\rm Pl}}},   
\end{equation}
$\langle \sigma | v |\rangle$ is the thermally averaged $\chi$-annihilation cross section and 
\begin{equation}
\eta_{\chi}^{\rm EQ} = 2 \left( \frac{m_\chi T}{2 \pi} \right)^{3/2} {\rm exp}\left(-\frac{m_\chi}{T}\right).
\end{equation}
The freeze out occurred when $\chi$'s were non-relativistic with $v \ll c$ and then $\langle \sigma | v |\rangle$ can be
written as
\begin{equation}
\sigma |v| = a + b v^{2} + \mathcal{O}(v^{4}),
\end{equation}
and we obtain the thermal-averaged cross section as
\begin{equation}
\langle \sigma \left( \chi\bar\chi \xrightarrow{S} f\bar f \right) | v |\rangle \simeq \frac{N_c^f c_s^4 m_\chi^2 \sqrt{1- \frac{4 m_f^2}{m_\chi^2}}}
{32 \pi \left( m_S^2 + m_\chi^2 - m_f^2 \right)^2},
\end{equation}
and
\begin{equation}
\langle \sigma \left( \chi\bar\chi \xrightarrow{V_\mu}  f\bar f \right) | v |\rangle \simeq \frac{N_c^f c_v^4 m_\chi^2 \sqrt{1- \frac{m_f^2}{m_\chi^2}}}
{8 \pi \left( m_V^2 + m_\chi^2 - m_f^2 \right)^2} \left\{1+\frac{m_\chi^2}{2 m_V^2} \left( 1 + \frac{m_f^4}{m_\chi^4}\right) 
\frac{m_\chi^4}{2 m_V^4} \left( 1 - \frac{m_f^4}{m_\chi^4}\right)   \right\},
\end{equation}
for the scalar and vector mediators respectively. $N_c^f = 3$ for quarks and 1 for leptons.

\begin{figure}[tbp]
\centering 
\includegraphics[width=.47\textwidth]{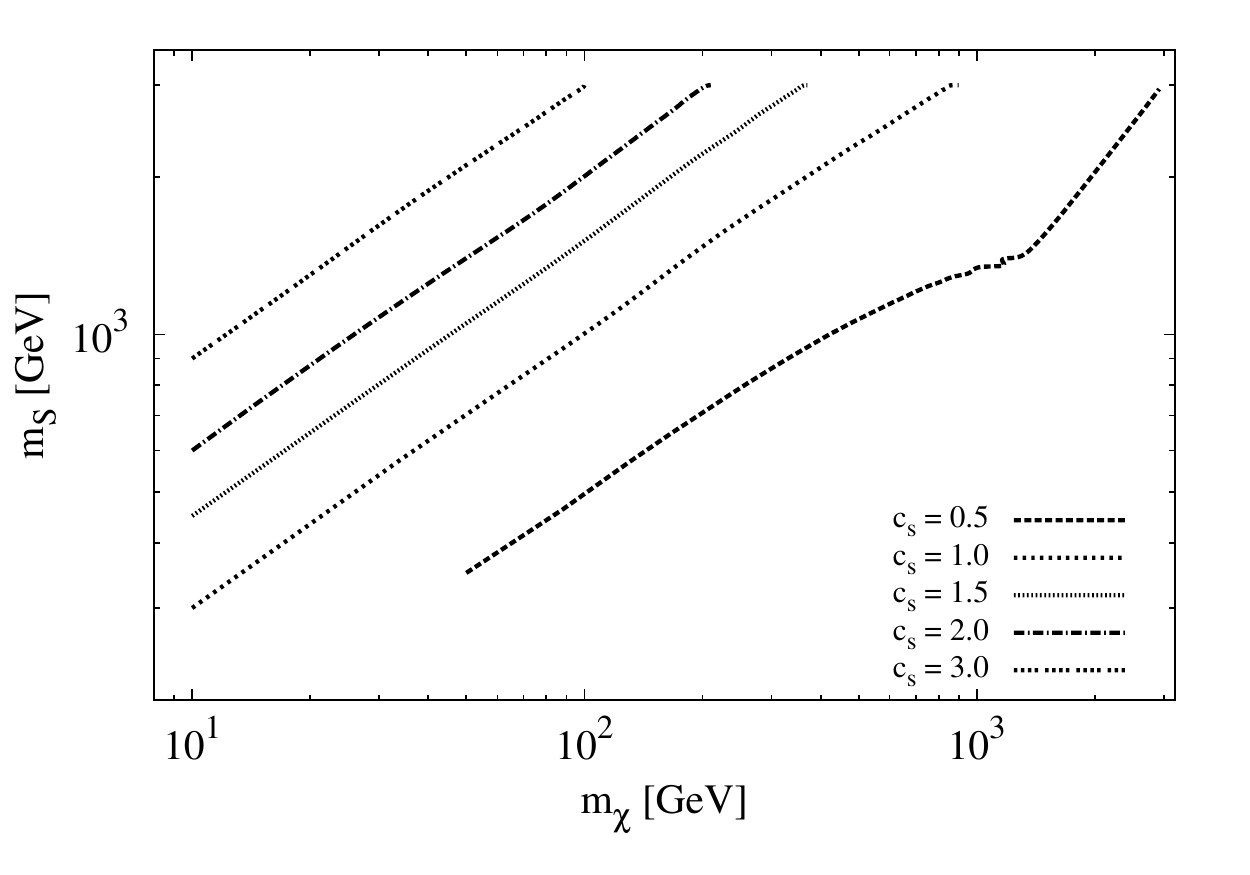}
\includegraphics[width=.47\textwidth]{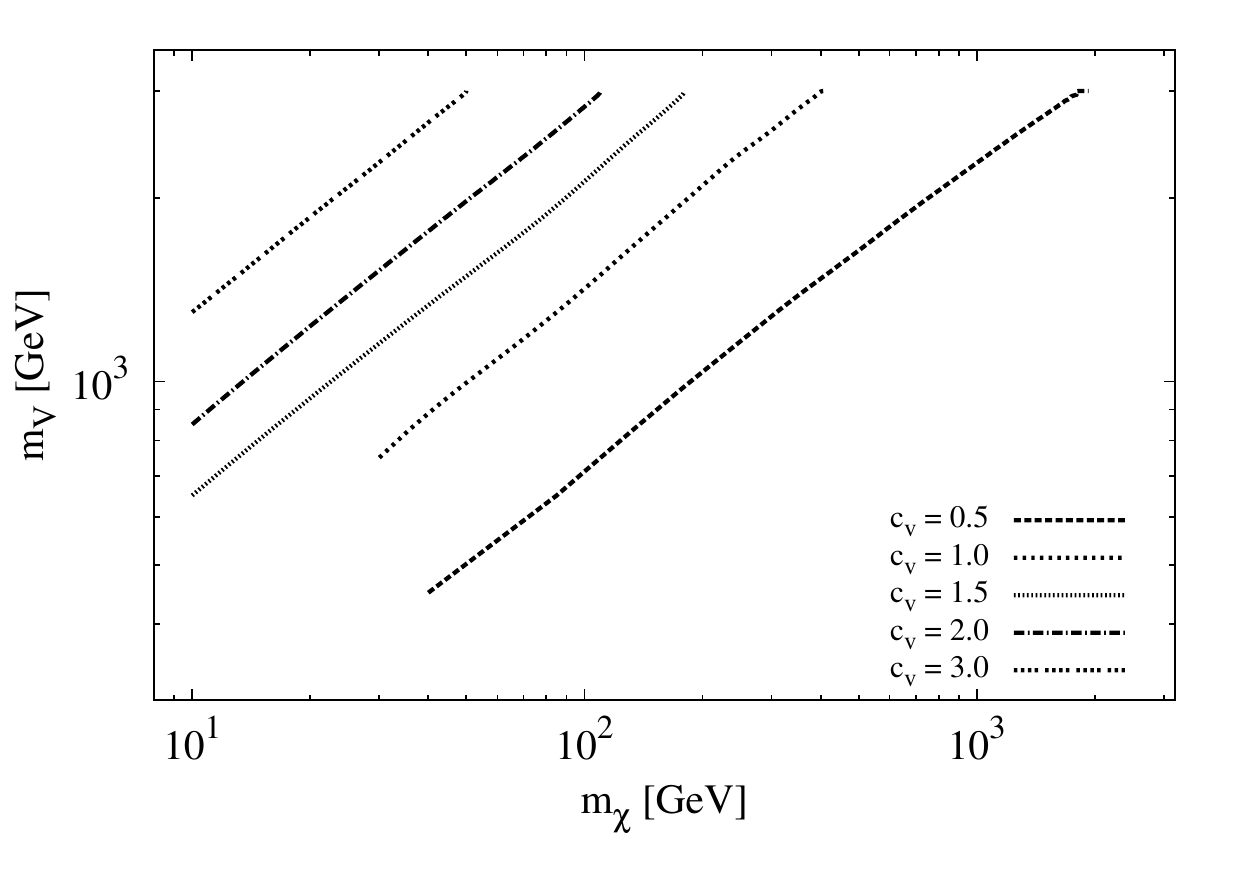}
\caption{\label{fig:i} Contour plots between the DM mass $m_\chi$ and the mediator mass $m_S/m_V$ under the
assumption that the DM $\chi$ saturates the observed relic density. The left and right panels are for the scalar and
vector mediators respectively. The contours are drawn for some representative values of the scalar and vector
couplings namely $c_{s/v} = $ 0.5, 1.0, 1.5, 2.0 and 3.0.}
\end{figure}
To calculate the relic density of the dark matter, we have implemented the scalar and vector $t$-channel mediator in 
\texttt{micrOMEGAs}~\cite{Belanger:2014vza} 
which numerically solves the Boltzmann equation by taking the full expression of the annihilation 
cross sections rather than the velocity expressions given here. Needed model file for \texttt{micrOMEGAs} are built using 
\texttt{FeynRules}~\cite{Alloul:2013bka}. 
In Figure~\ref{fig:i} we show the contours of the relic density within 5$\sigma$ of the observed 
value~\cite{Ade:2015xua} for fixed coupling values for the scalar and vector mediators respectively.

\begin{figure}[tbp]
\centering 
\includegraphics[width=.47\textwidth]{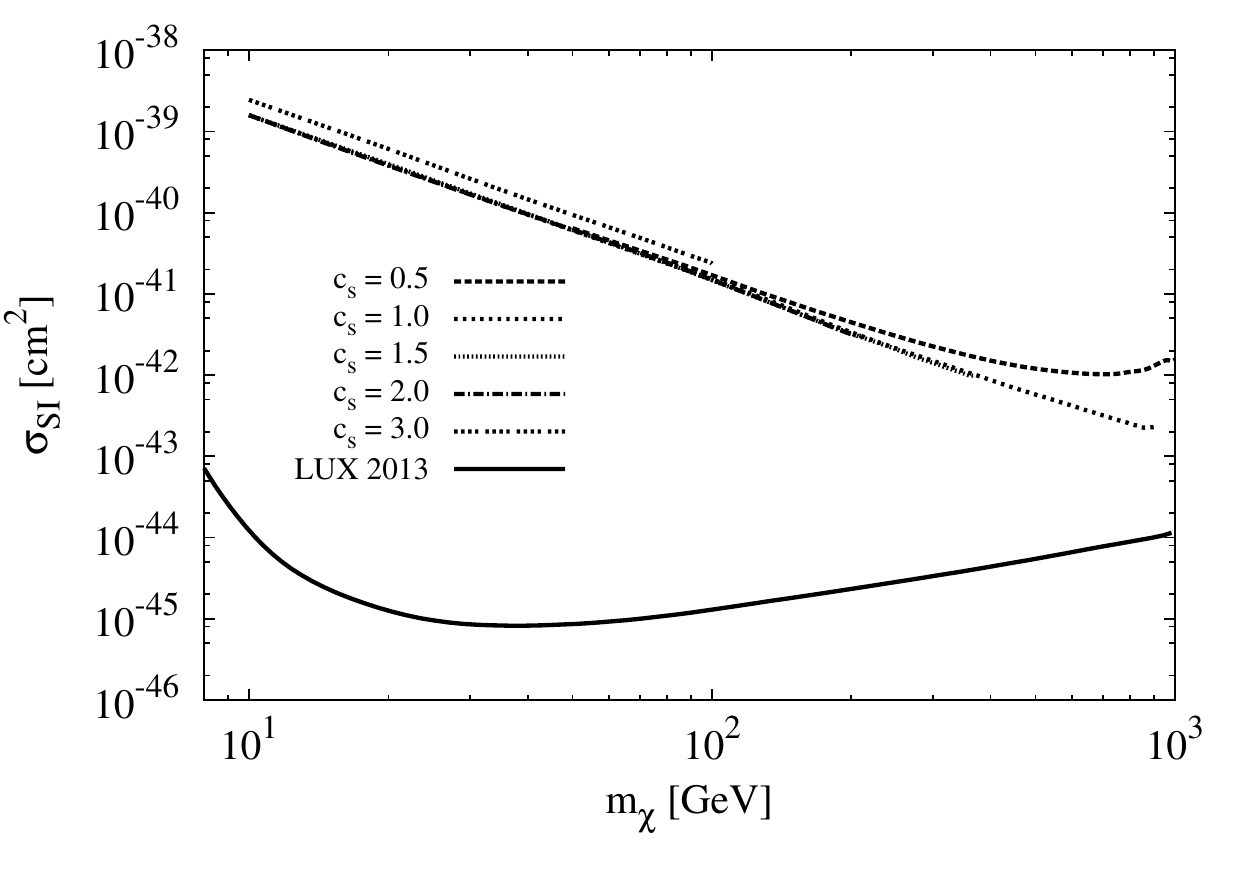}
\includegraphics[width=.47\textwidth]{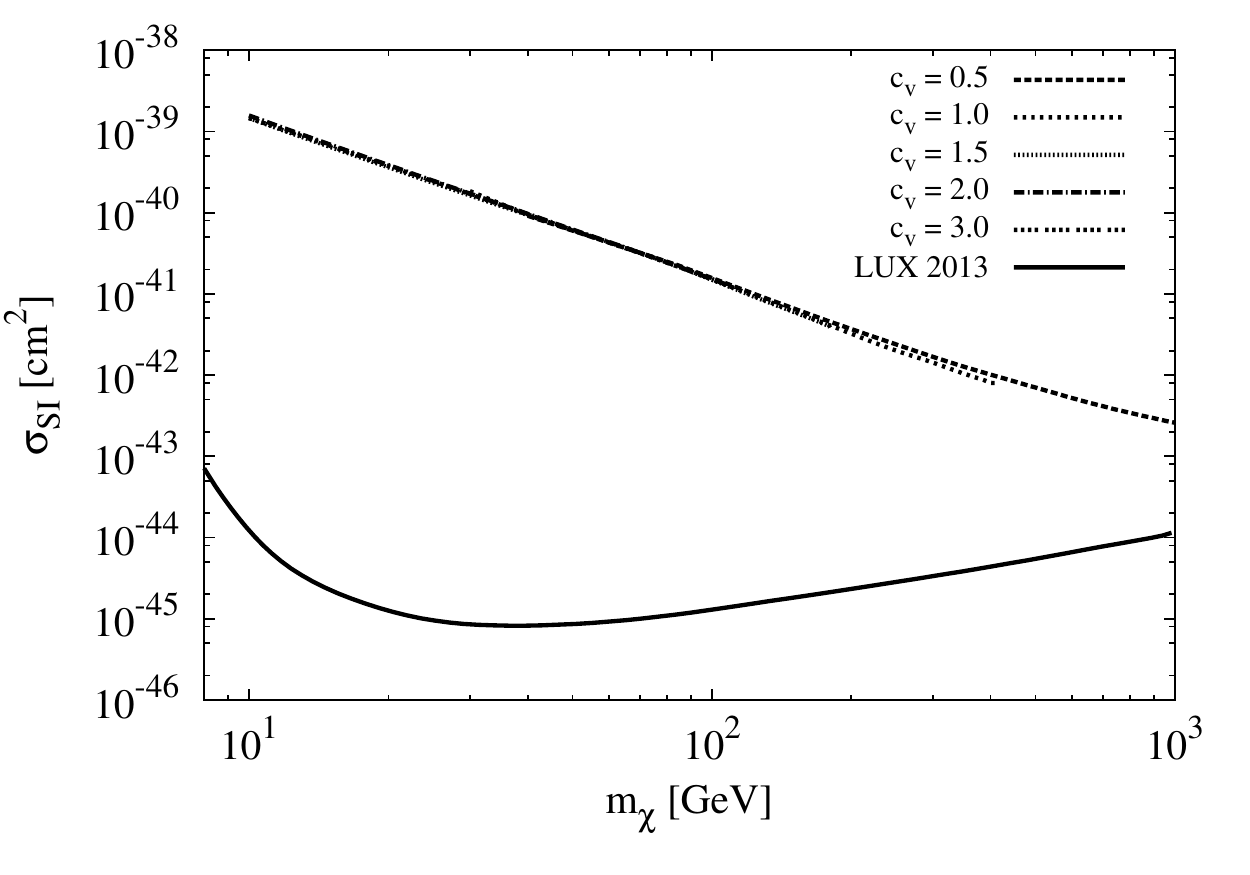} \\
\includegraphics[width=.47\textwidth]{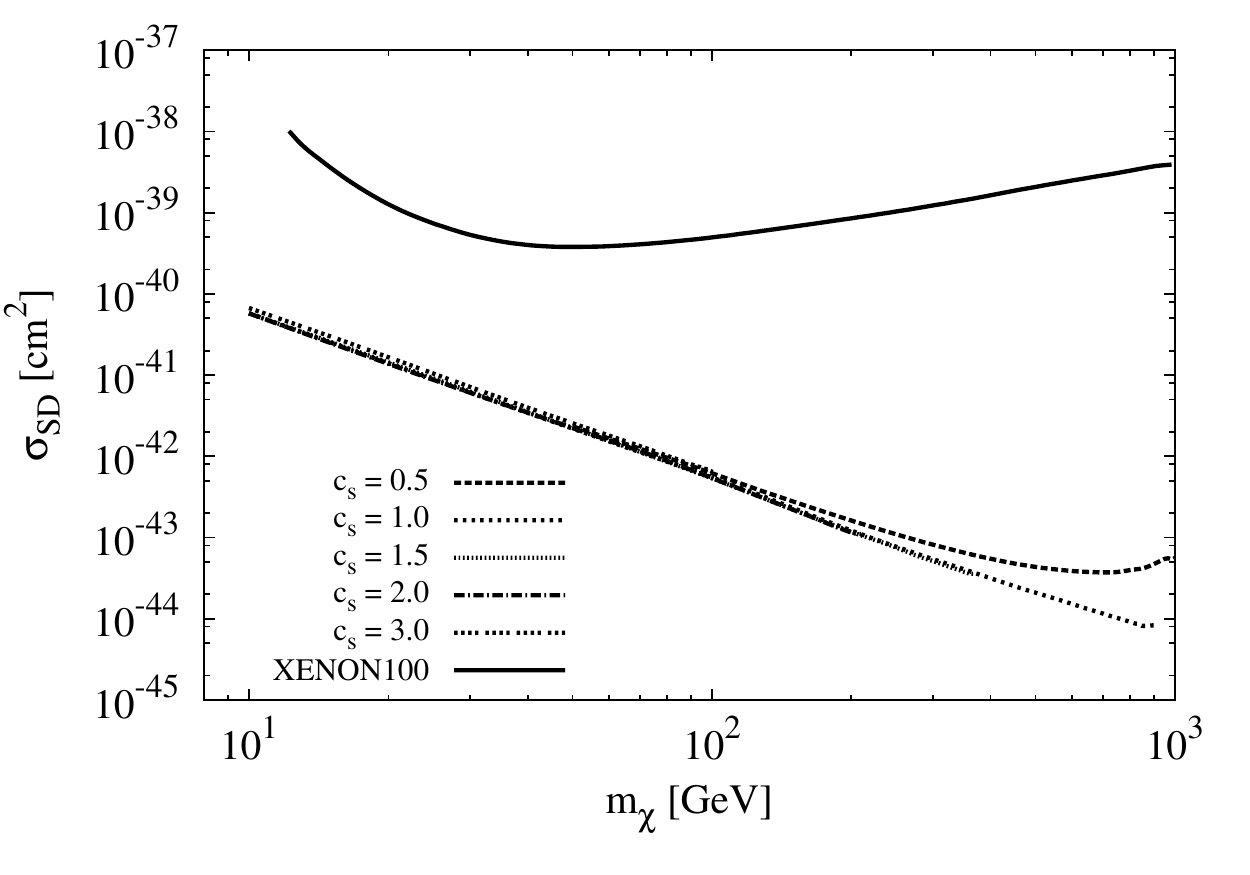}
\includegraphics[width=.47\textwidth]{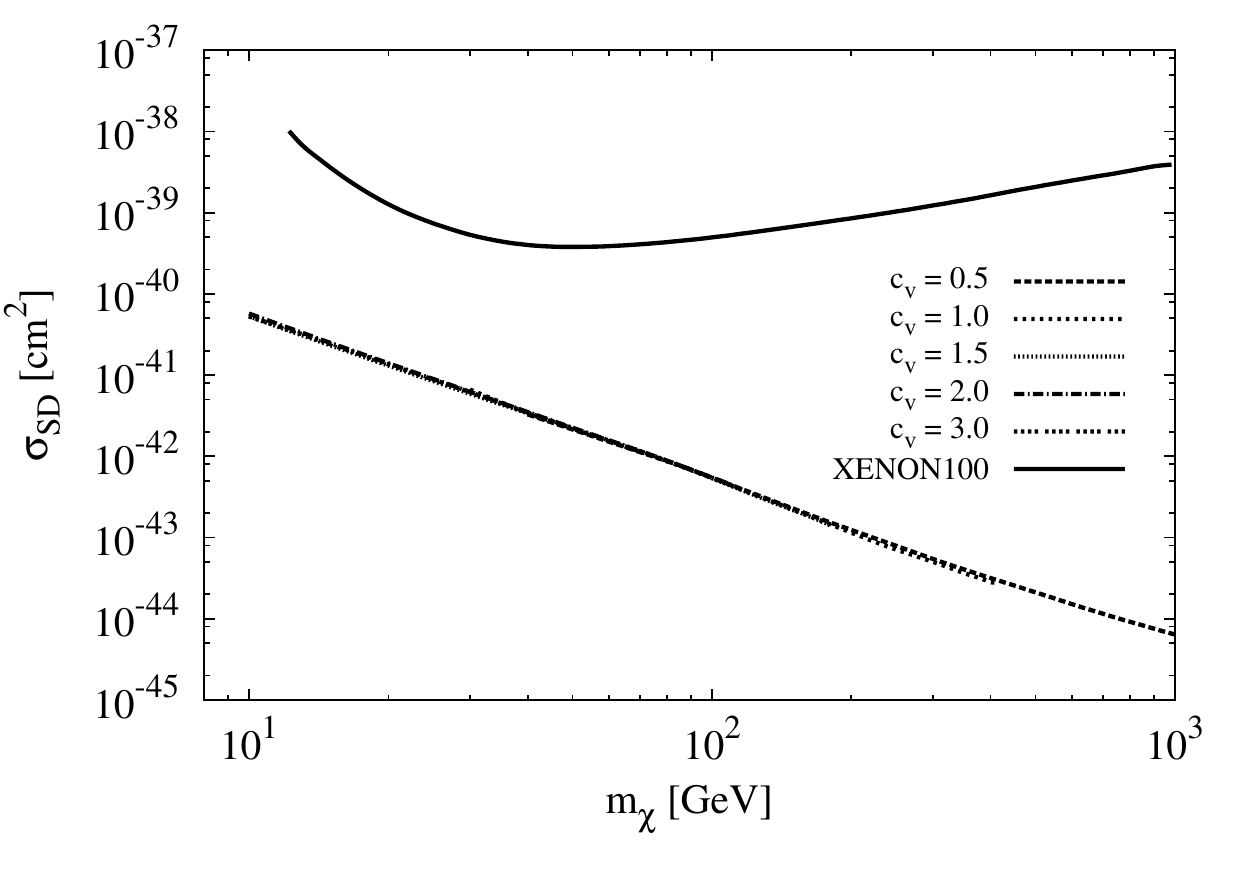} 
\caption{\label{fig:ii} The spin-independent DM-nucleon scattering cross section $\sigma^{\rm SI}$ (top panel) and
the spin-dependent cross section $\sigma^{\rm SD}$ (bottom panel). The LUX bound for 2013 data~\cite{Akerib:2013tjd}
 for $\sigma^{\rm SI}$
and the XENON100 neutron bounds~\cite{Aprile:2013doa}
for $\sigma^{\rm SD}$ are also shown. The left and the right panels are for the
scalar and vector mediators respectively. All cross sections are drawn for the parameter values $m_\chi, m_S/m_V$
and $c_{s,v}$ consistent with the relic density constraints. The LUX bound for $\sigma^{\rm SI}$ practically rules
out the entire parameter space allowed by the observed relic density both for the case of scalar as well as vector
mediators. }
\end{figure}

\subsection{Direct detection}
\label{dir}    
In order to obtain constraints from dark matter direct detection experiments, we need to calculate the nucleon-DM elastic
cross section. In the non-relativistic limit the elastic nucleon-DM cross sections can be easily calculated and we obtain both the 
spin-independent and subdominant spin-dependent contributions. For the case of scalar mediators, the cross sections are
given as
\begin{align}
\sigma^{\rm SI} \left(\chi N \xrightarrow{S} \chi N \right) &= \frac{c_s^4 \mu^2}{64 \pi \left( m_S^2 - m_\chi^2 \right)^2} f_N^2 \label{ssis}\\
\sigma^{\rm SD} \left(\chi N \xrightarrow{S} \chi N \right) &= \frac{c_s^4 \mu^2}{64 \pi \left( m_S^2 - m_\chi^2 \right)^2} \left( \Delta q^N \right)^2 \label{ssds}.  
\end{align} 
The corresponding cross sections for the vector-mediator are
\begin{align}
\sigma^{\rm SI} \left(\chi N \xrightarrow{V_\mu} \chi N \right) &= \frac{c_v^4 \mu^2}{16 \pi m_V^4} f_N^2 \label{ssiv}\\
\sigma^{\rm SD} \left(\chi N \xrightarrow{V_\mu} \chi N \right) &= \frac{9 c_v^4 \mu^2}{16 \pi \left( m_V^2 - m_\chi^2 \right)^2} \left( \Delta q^N \right)^2 \label{ssdv},  
\end{align} 
where $\mu=\frac{\mu_\chi \mu_N}{\mu_\chi + \mu_N}$ is the reduced mass and 
\begin{align}
&f_p = f_n = f_N = 3 \\
&\Delta u^{(p)} = \Delta d^{(n)} = 0.84 \pm 0.02 \\ 
&\Delta d^{(p)} = \Delta u^{(n)} = - 0.43 \pm 0.02 \\
&\Delta s^{(p)} = \Delta s^{(n)} =  - 0.09 \pm 0.02 \\ 
\end{align}
In Figure~\ref{fig:ii} we show the predictions for the spin-independent $\sigma^{\rm SI}$ and spin-dependent $\sigma^{SD}$ 
cross sections as a
function of DM mass $m_\chi$ for scalar and vector mediators respectively. The corresponding experimental bounds from 
LUX~\cite{Akerib:2013tjd} and
XENON100~\cite{Aprile:2013doa}
are also displayed. For all the cross sections shown the scalar and vector mediator masses $m_S$ and $m_V$ are set 
to give the observed relic density within $5\sigma$ for all values of $m_\chi$ and couplings $c_s$ and $c_v$. A noticeable feature
of the DM-nucleon elastic cross sections is their almost total independence with respect to the scalar and vector couplings $c_s$ and
$c_v$ for parameters consistent with the observed relic density constraints. This seems to happen because an increase in coupling
is accompanied by a corresponding increase in the mediator mass roughly by the same ratio as can be ascertained from the relic
density contours in Figure~\ref{fig:i}.     

\begin{figure}[tbp]
\centering 
\includegraphics[width=.47\textwidth]{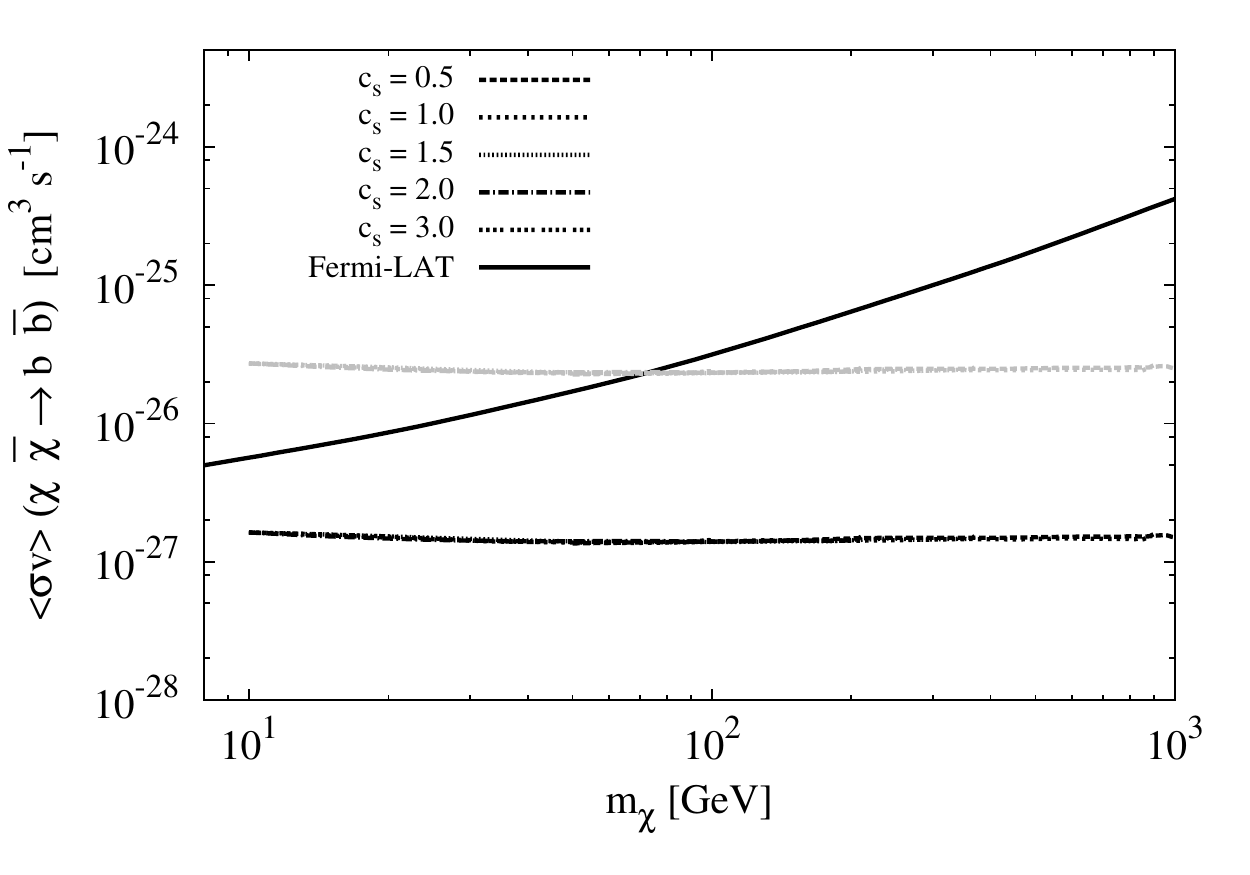}
\includegraphics[width=.47\textwidth]{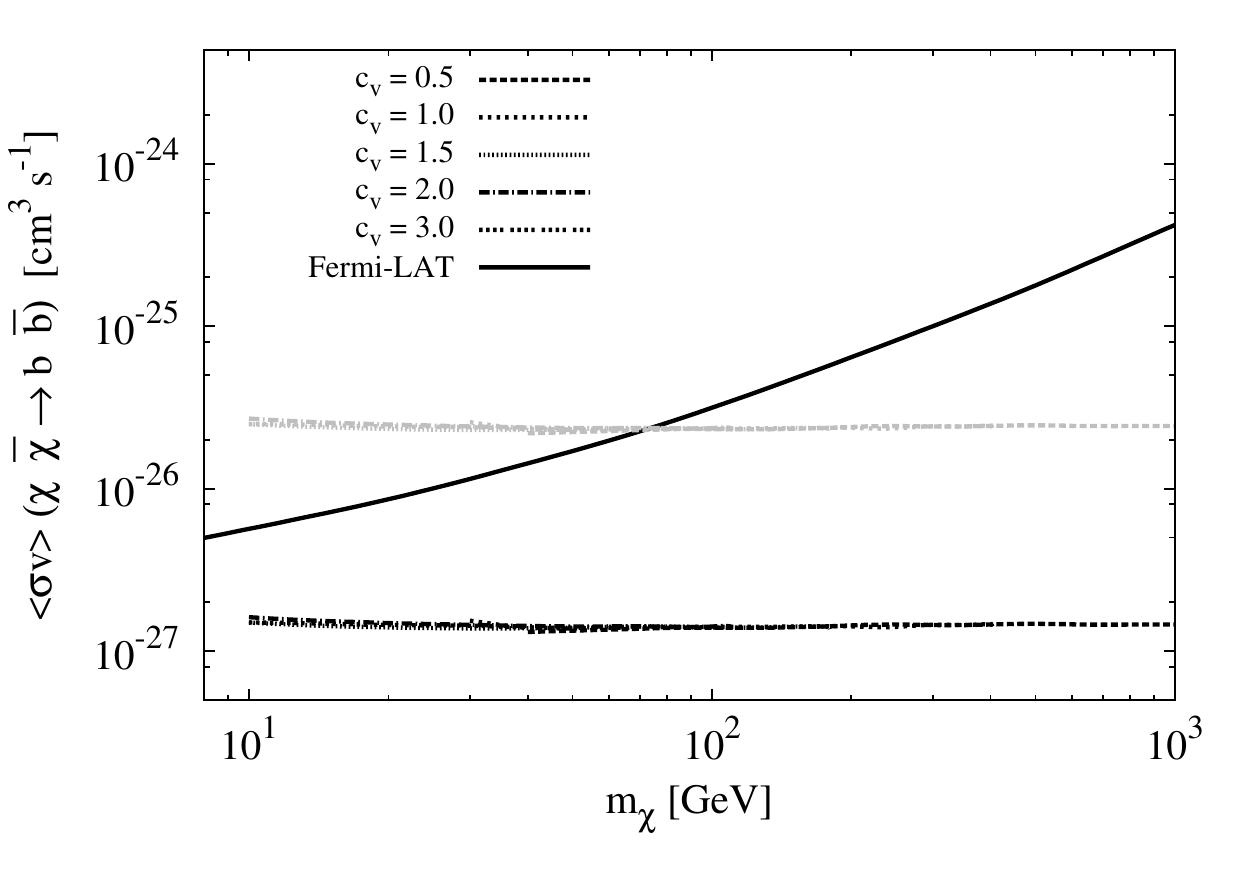}
\caption{\label{fig:iii} The prediction for the DM $\chi\bar \chi$ annihilation rate into $b\bar b$ (total in grey colour) 
as a function of DM mass $m_\chi$. The mediator mass $m_S$ and $m_V$ is fixed from the observed relic density. 
The left and the right panels corresponds to the scalar 
and vector mediators respectively. Bounds for the Fermi-LAT experiments are also shown.}
\end{figure}

\subsection{Indirect detection}
\label{indir}
DM annihilation in the universe into SM particles would result in cosmic ray fluxes which can be observed by dedicated detectors. The Fermi Large Area Telescope (LAT) collaborations~\cite{Ackermann:2015zua}
has produced constraints into DM annihilation cross sections into final states
like $e^+e^-,$ $\mu^+\mu^-,$ $\tau^+\tau^-,$ $u\bar u,$ $b\bar b,$ $W^+W^-$ etc.

In Figure~\ref{fig:iii} we show the prediction for DM annihilation into $b\bar b$ for some representative values of the scalar and vector 
couplings as a function of $m_\chi$. The predictions shown here for different values of couplings and DM mass $m_\chi$ are consistent 
with the observed relic density. We have also shown the bounds from the Fermi-LAT experiments. We observe similar features as
seen in DM-nucleon scattering cross sections. The Fermi-LAT observations do not put any strong bounds on the parameters.       

\begin{figure}[tbp]
\centering 
\includegraphics[width=.47\textwidth]{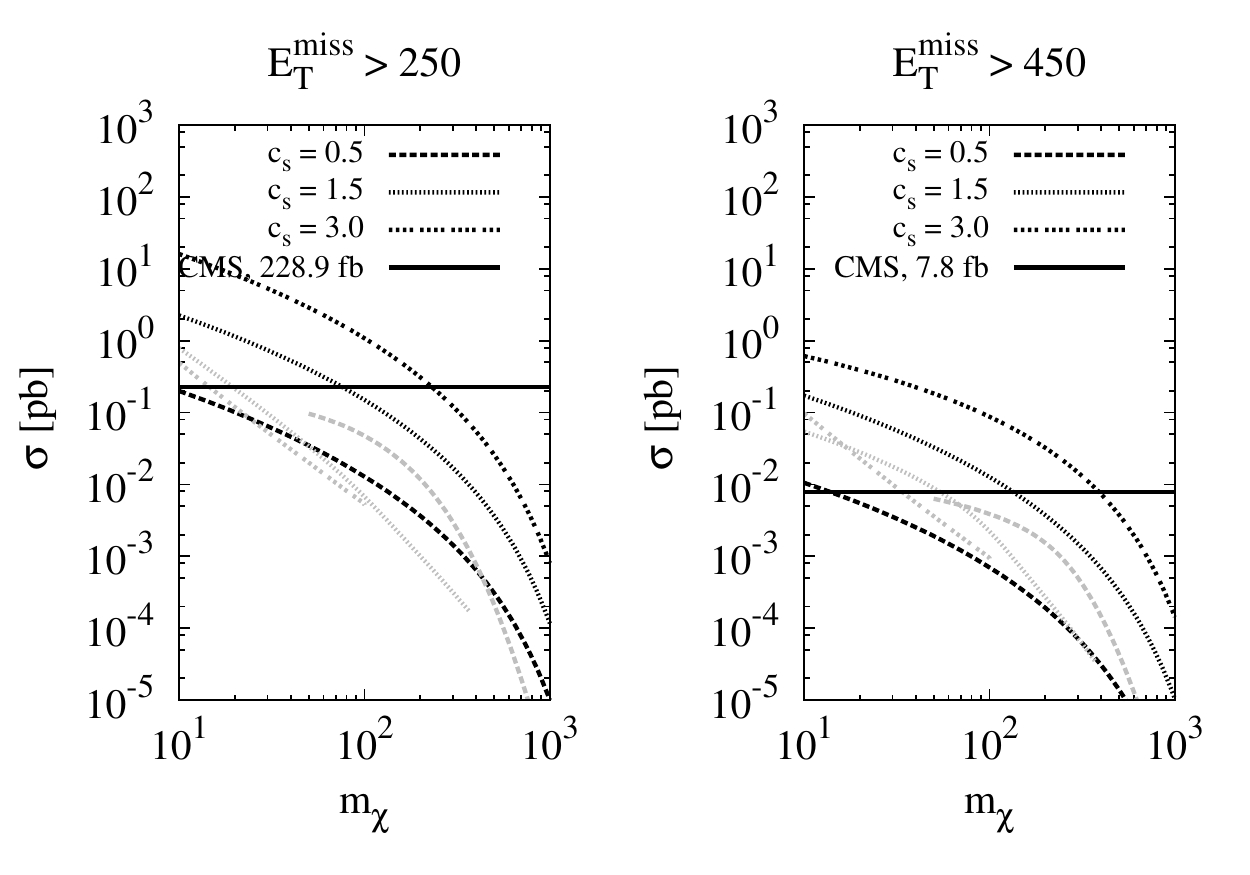}
\includegraphics[width=.47\textwidth]{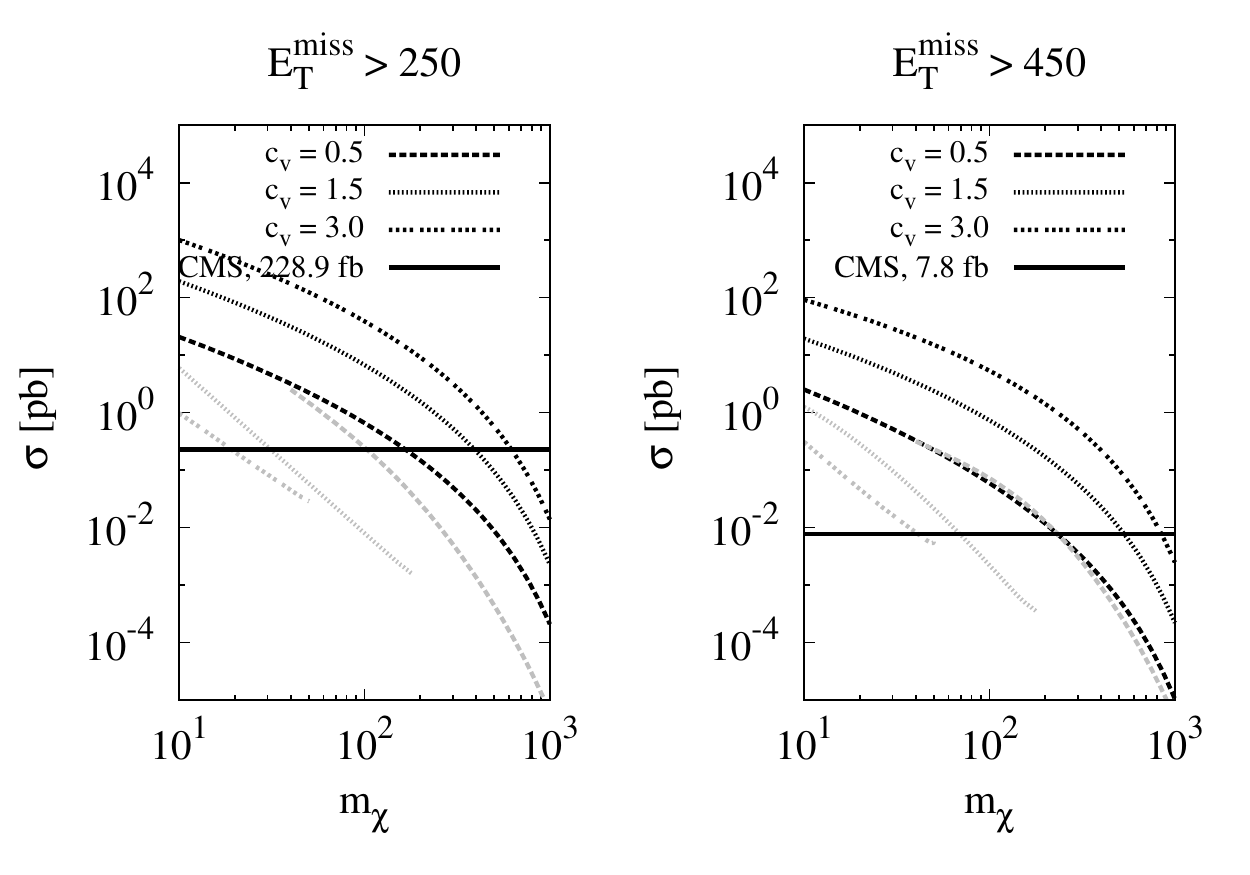}
\caption{\label{fig:iv} The monojet cross section in (pb) at the LHC $\sqrt s = 8$~TeV for two cases (i) $E_T^{\rm miss} > 250$~GeV
and (ii) $E_T^{\rm miss} > 450$~GeV. The cross sections are obtained by scanning the ($m_\chi, m_S/m_V$) parameter
space for benchmark couplings shown in the figures. The grey curves corresponds to cross sections obtained for the
parameter values consistent with the observed relic density. The allowed parameter space for fermionic spin-1/2 DM
lies below the CMS bound for (i) $\sigma_{\rm monojet} = 228.9$ fb and (ii) $\sigma_{\rm monojet} = 7.8$ fb as
explained in the text. We observe that low DM mass values ($m_\chi \lesssim 100$~GeV) are not allowed by the CMS
monojet data for all values of coupling considered.}
\end{figure}

\subsection{Collider bounds}
\label{collider}    
The $t$-channel mediator model has scalar and vector mediator particles which carry colour and thus couple to quarks and SM
singlet dark matter particle $\chi$. They can thus be singly produced in association with a DM particle or pair produced if they are
light enough at the LHC through the QCD and DM exchange processes. These processes will contribute to the monojet and dijet
signals with missing energy ($E_T^{\rm miss}$). Since the scalar and vector mediators also carry SU(2)$_L$ and U(1) charges,
we will also have mono-W and mono-photon plus missing energy signals with distinct signature that can be searched in 
dedicated searches. These processes proceed through electroweak interactions and are expected to be subdominant.
Mono-W signature in an $s$-channel vector mediator simplified model has been studied in~\cite{Haisch:2016usn}. 
Here we will confine ourselves to QCD processes.   

For the monojet events $qg \to q \chi \bar \chi$ process are the dominant process because of the large parton distribution probability
of the gluon as compared to the quark or anti-quark in the proton. It has been emphasised by the authors of the simplified dark model
document~\cite{Abdallah:2014hon} 
that the dominance of the associated production channels is a distinct feature of $t$-channel models. The most
stringent limit is from the CMS collaborations which has used 8 TeV data based on an integrated luminosity of 
19.7 $\rm fb^{-1}$~\cite{Askew:2014kqa, Khachatryan:2014rra}.
The CMS data has been used by the authors of Ref.~\cite{An:2013xka} 
to put bounds on the coupling as a function of the mediator and DM mass
for the case of scalar $t$-channel mediator model. It may be mentioned that the authors of 
Ref.~\cite{Abdallah:2014hon} used the ATLAS and CMS
collaboration data~\cite{CMS:2013gea, TheATLAScollaboration:2013fha}
on the limit to squark pair production cross section to constraint $t$-channel mediator model from 
dijet plus missing energy signals.  

In our study here we confine ourselves to constraints arising from the monojet signals.
For this study of monojet constraints, we use the parameter space ($m_\chi, m_{S/V}$)
for different values of coupling $c_s/c_v$ consistent with the observed DM relic density. To obtain the cross section for monojet events,
we generate parton level events for the process $p p \to \chi \bar \chi j$ using \texttt{MadGraph}~\cite{Alwall:2014hca} 
where the model file for the
Lagrangian is obtained from \texttt{FeynRules}~\cite{Alloul:2013bka} 
and we use \texttt{CTEQ6l1} parton distribution function for five flavour quarks
in the initial state. We employ the usual cuts and the cross sections are calculated here to put bounds on the parameters of 
the model by requiring (i) $E_T^{\rm miss} > 250$~GeV and (ii) $E_T^{\rm miss} > 450$~GeV for which the CMS result excludes new
contribution to the monojet cross section exceeding 228.9 fb and 7.8 fb at 95\% C.L respectively. The resulting monojet cross section 
for the scalar and vector mediators are shown in Figure~\ref{fig:iv}
as function of $m_\chi$ for the values of the mediator mass $m_{S/V}$ consistent with the observed relic density. The results are
displayed for some representative values of the couplings. We find that the bounds from monojet searches are much weaker in 
comparison to the bounds from the direct searches by LUX based on spin-independent DM-nucleon scattering. The spin-dependent
elastic DM-nucleon scattering data by XENON100 observations on the other hand allows all values of model parameters expressed 
in this study which are consistent with the observed relic density whereas the monojet searches put strong constraints on the
parameters. In addition we find that low DM mass ($m_\chi \lesssim 100$~GeV) is not allowed by the CMS monojet searches for all 
values of couplings considered. This is true for the scalar as well as vector mediators.

\begin{figure}[tbp]
\centering 
\includegraphics[width=.47\textwidth]{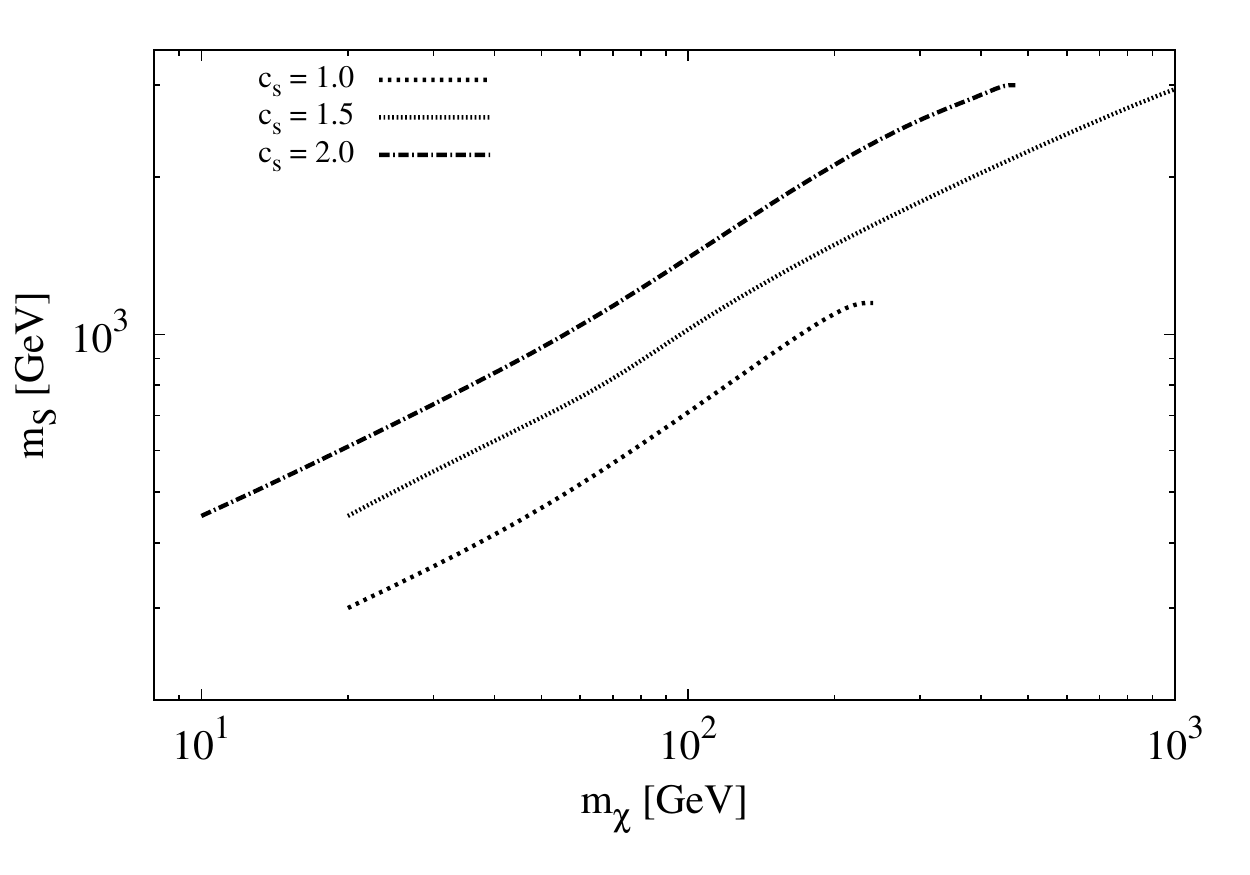}
\includegraphics[width=.47\textwidth]{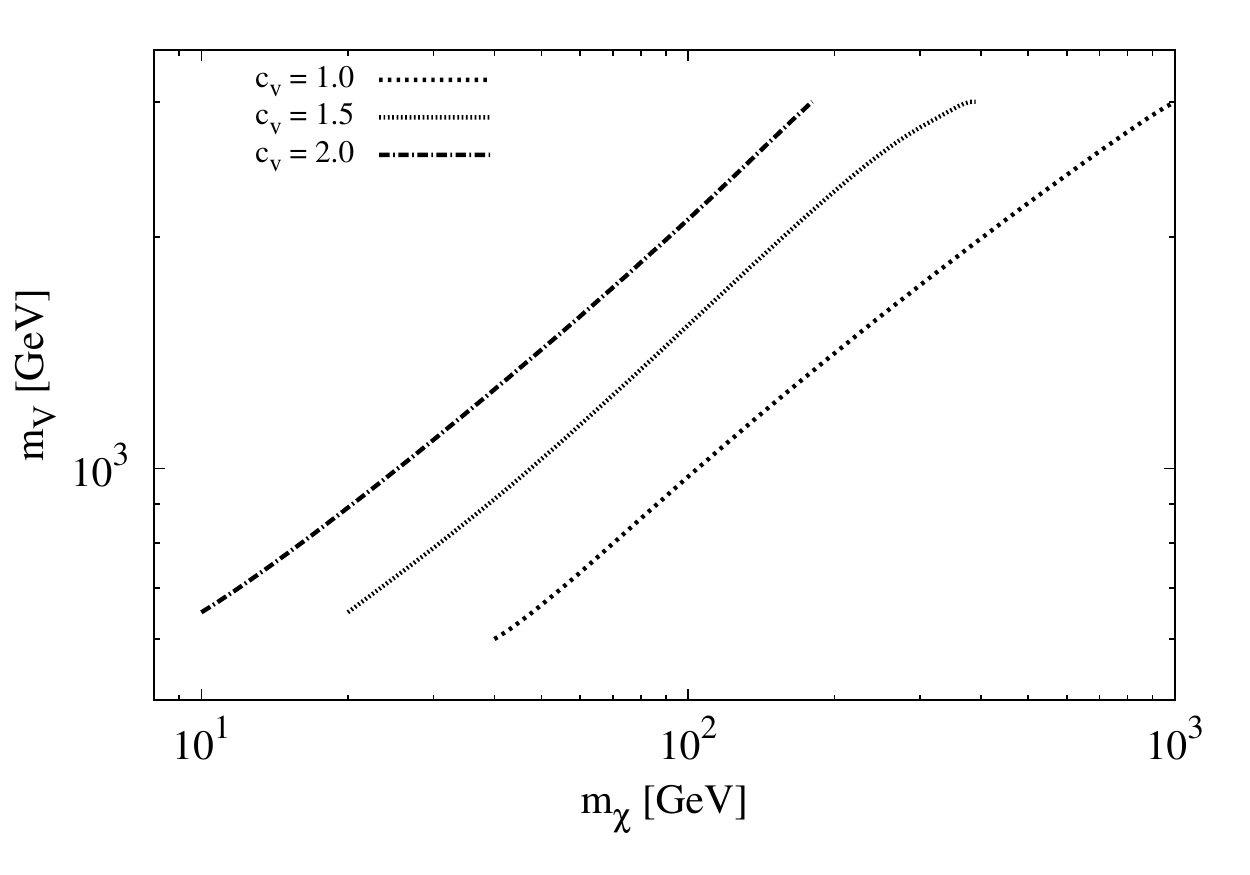} \\
\includegraphics[width=.47\textwidth]{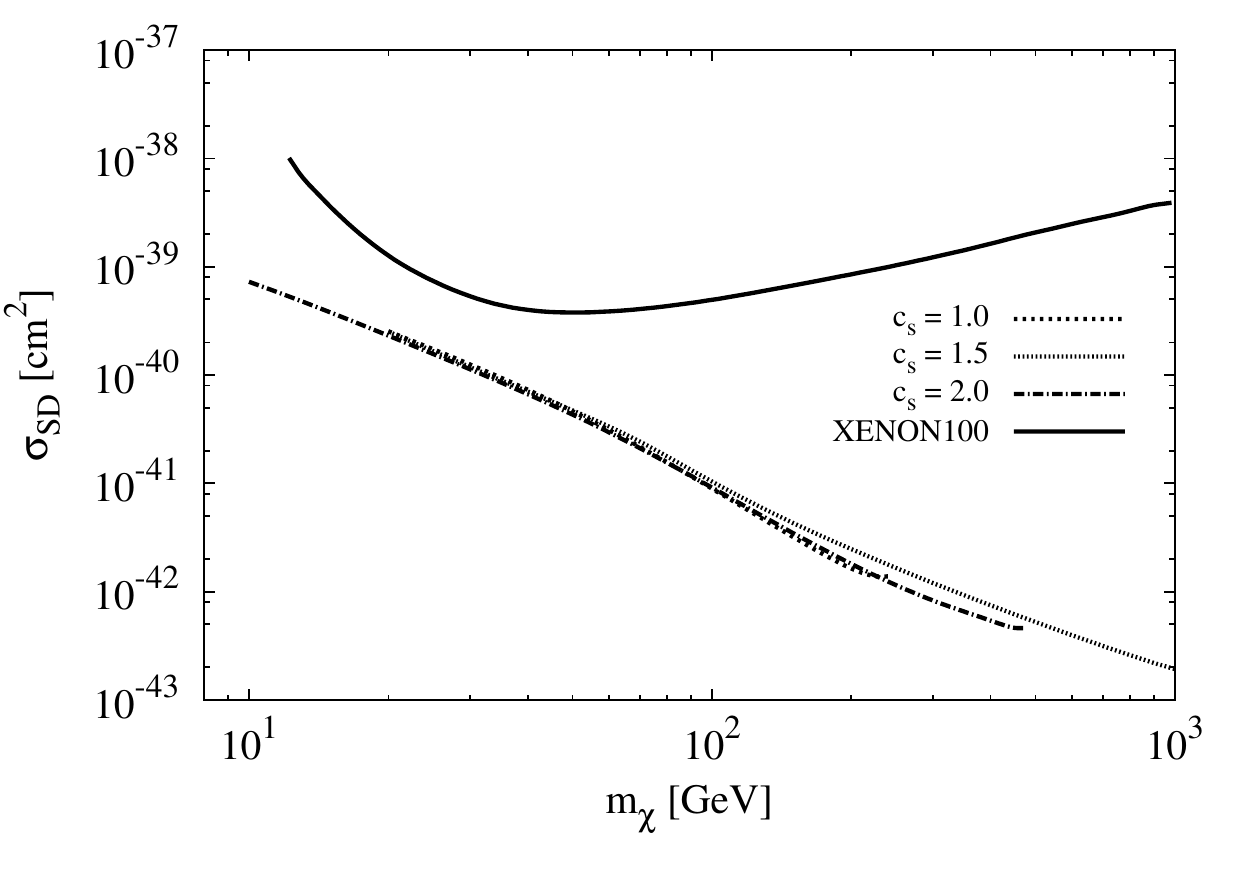}
\includegraphics[width=.47\textwidth]{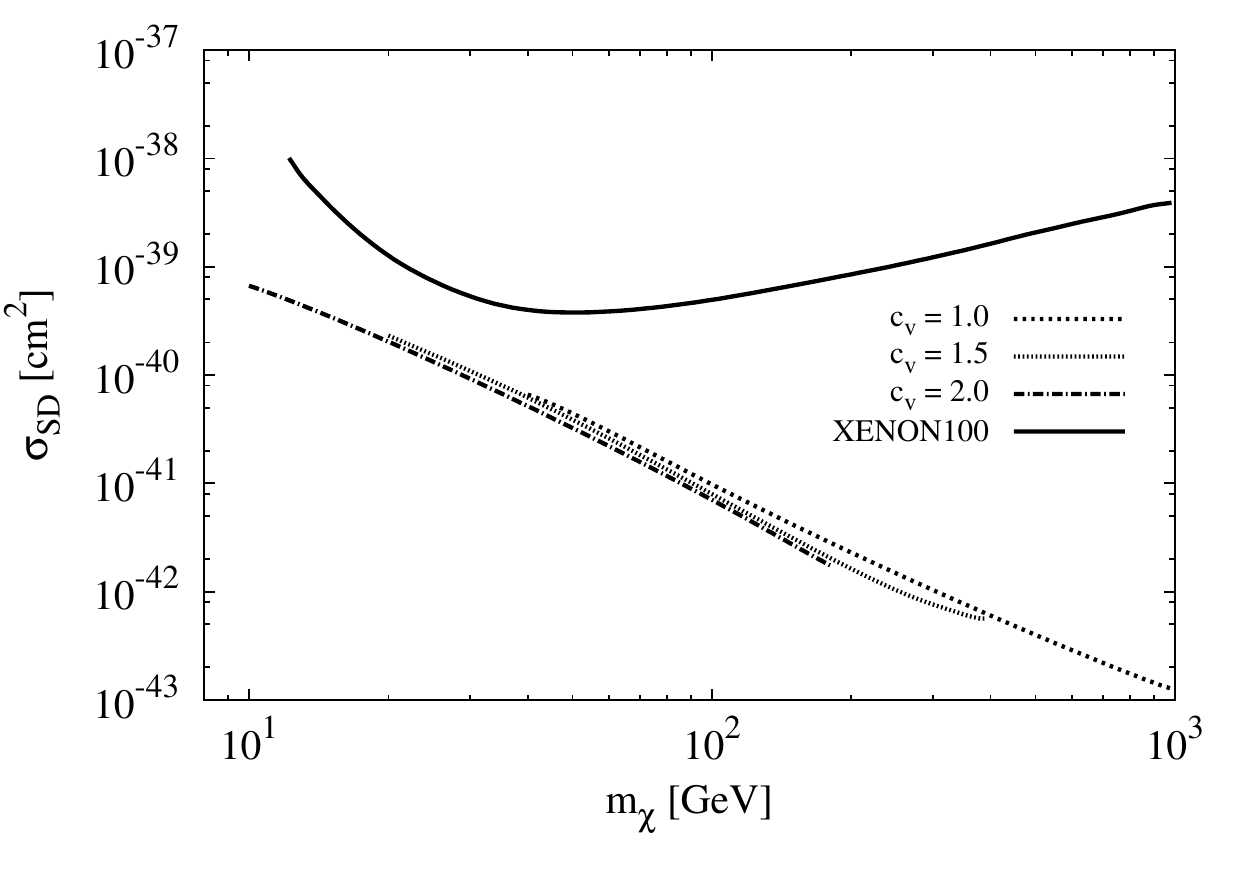} \\
\includegraphics[width=.47\textwidth]{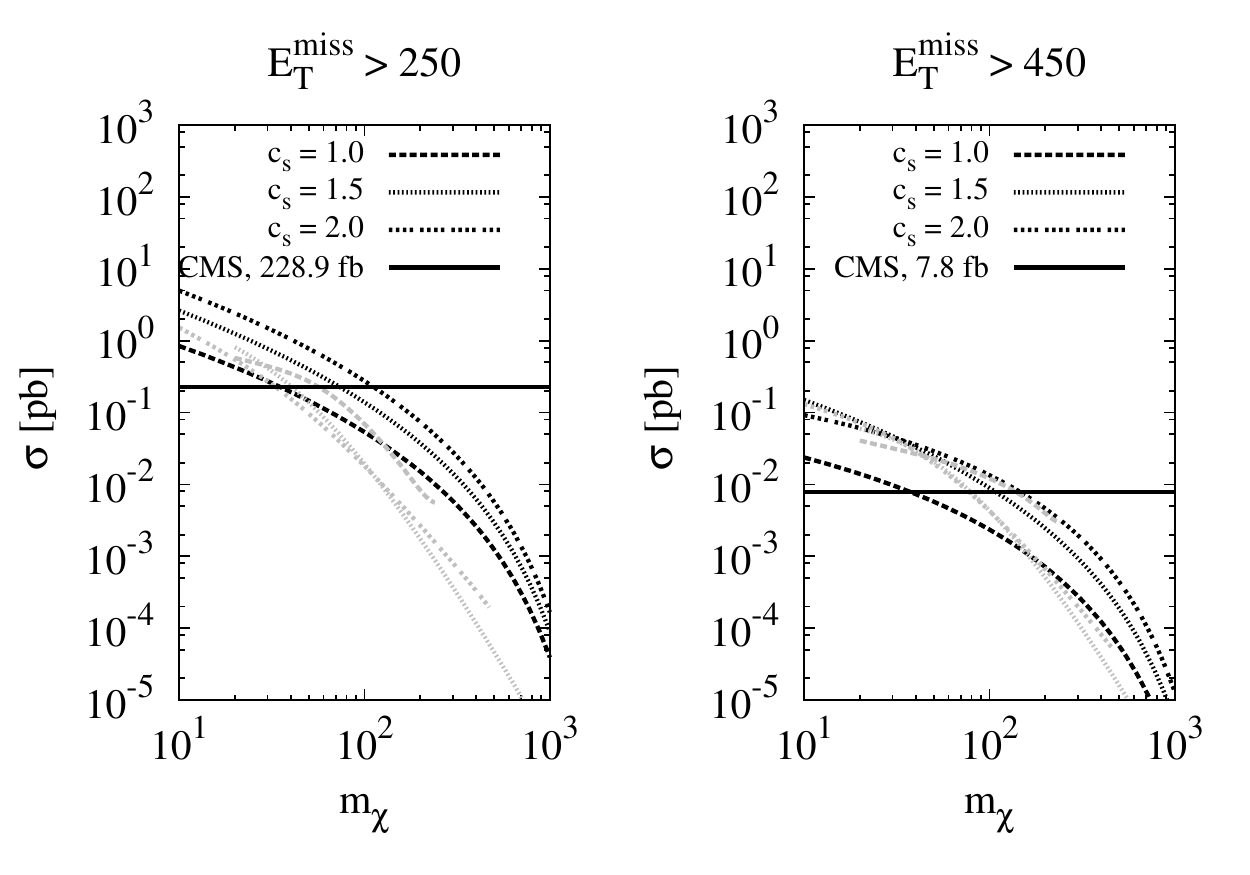}
\includegraphics[width=.47\textwidth]{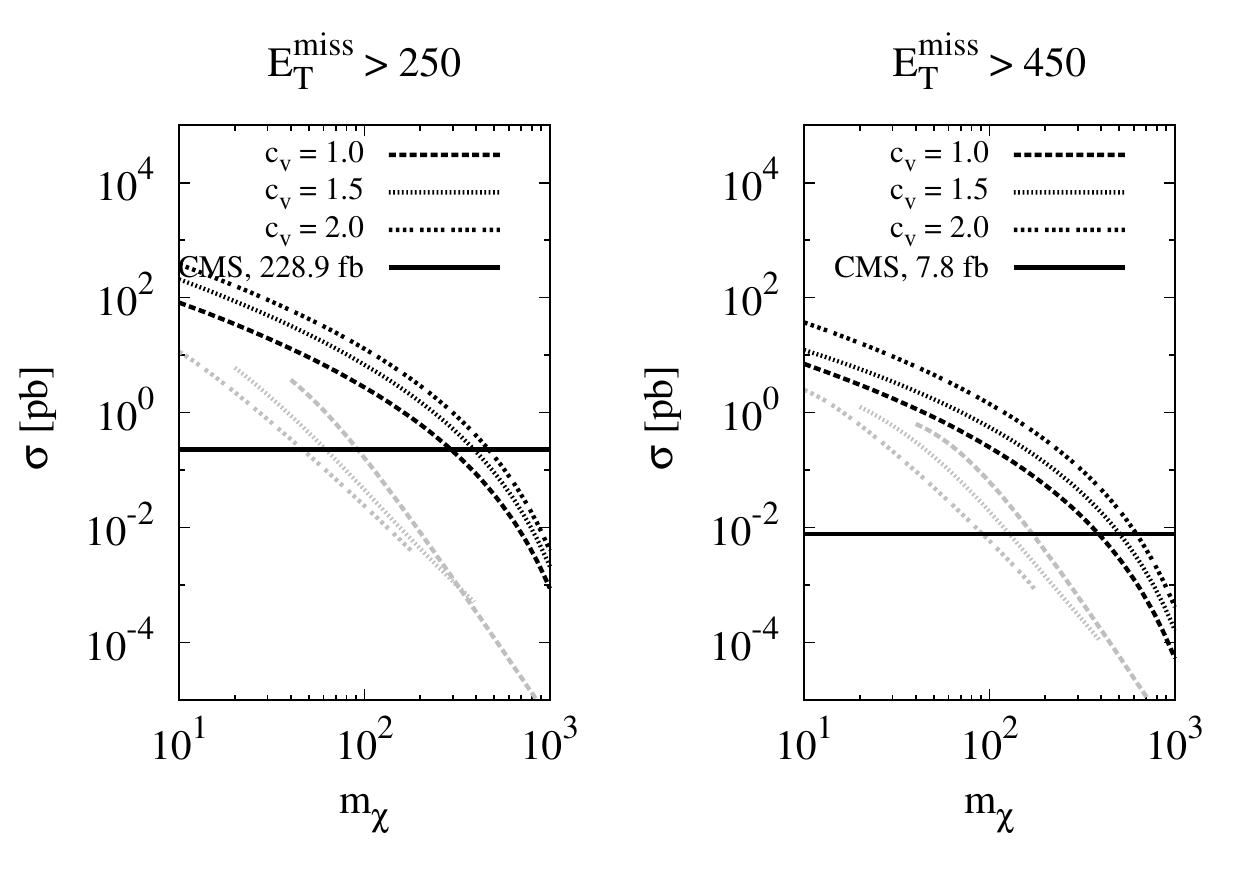}
\caption{\label{fig:v} Results for the Majorana DM. Top panel: Contour plots between $m_\chi$ and mediator mass
$m_S/m_V$ under the assumption that DM $\chi$ saturates the observed relic density. Middle panel: The spin-dependent
DM-nucleon cross sections along with the XENON100 bounds for $\sigma^{\rm SD}$. All cross sections are drawn for 
parameter values consistent with the relic density constraints. Bottom panel: The monojet cross section in (pb) at the LHC 
$\sqrt s = 8$~TeV for two cases (i) $E_T^{\rm miss} > 250$~GeV and (ii) $E_T^{\rm miss} > 450$~GeV. The cross sections
are obtained as in Figure~\ref{fig:iv}. The curves in grey corresponds to $p p \to \chi \chi + j$ cross sections consistent
with the observed relic density. The allowed parameter space for Majorana DM lies below the CMS bound for 
$\sigma_{\rm monojet} = 228.9 \,(7.8)$ fb corresponding to $E_T^{\rm miss} > 250 \,(450)$~GeV. The left and the right panels
are for the scalar and vector mediators respectively. We observe that for Majorana DM strong constraints on the minimum
DM mass exist.}
\end{figure}

\section{Summary and discussions}
\label{summ}    
In this paper we have considered a fermionic DM particle in renormalizable SM gauge interactions in a simple $t$-channel model.
The DM particle interactions with the SM fermions is through the exchange of scalar and vector mediators which carry colour. We
have imposed MFV by considering universal couplings with all generations of mediators to be equal. The main observations are
the following:
\begin{itemize}
\item[(1)] For the scalar as well as vector mediators the entire parameter space ($m_\chi, m_S/ m_V$) consistent with the
observed relic density is ruled out by spin-independent direct DM-nucleon elastic scattering bounds given by LUX 2013
data for all the benchmark couplings considered. The spin-dependent DM-nucleon scattering bounds from XENON100 data
however, allows all the parameter values consistent with the observed relic density. This is in agreement with the findings of
Ref.~\cite{Abdallah:2014hon}.
\item[(2)] In case of DM $\chi$ being a Majorana fermion, the spin-independent $\chi$-nucleon scattering cross section
vanishes at the tree level. The spin-independent operators can however be generated at the one or two loop level~\cite{Drees:1993bu},
but are considerably suppressed and the dominant signal remains the spin-dependent $\chi$-nucleon scattering cross
section which for the Majorana fermion is four times that for the Dirac-fermion in Equations~\ref{ssds} 
and~\ref{ssdv}~\cite{Abdallah:2014hon}.
Recently singlet Majorana DM in the context of simplified models and dimension six EFT models has been considered 
in Ref.~\cite{Matsumoto:2016hbs}.

Thus if the DM is a Majorana fermion, all bounds from direct, indirect and collider observations can be seen to be consistent
with values of $m_\chi, m_S/m_V$ required to explain the observed relic density. In Figure~\ref{fig:v} we have the results for the
Majorana DM. In the top panel we have shown the contour plots between the Majorana DM mass $m_\chi$ and the 
mediator mass $m_S/m_V$ for some representative values of the couplings under assumption that $\chi$ saturates the
observed relic density. In the middle panel we have shown the corresponding spin-dependent DM-nucleon scattering cross
sections along with the XENON100 bounds. In the bottom panel monojet cross sections at the LHC for the process 
$p p \xrightarrow{S/V} \chi \chi + j$ in (pb) for two cuts on $E_T^{\rm miss}$ are displayed. The grey curves corresponds to
cross sections obtained for the parameter values consistent with the observed relic density. The allowed parameter space
for Majorana fermions lie below the CMS bound for $\sigma_{\rm monojet} = 228.9 (7.8)$ fb for $E_T^{\rm miss} > 250 (450)$~GeV.
The left and the right panels are for scalar and vector mediators respectively. We observe that for Majorana DM, strong constraints are
obtained on the minimum DM mass for all values of the coupling considered. The bounds from the monojet signal are
strongest for the case of vector mediator $viz.$ DM mass $\gtrsim 200$~GeV is consistent with relic density constraints.
\item[(3)] In the literature a class of simplified models has been considered where the DM $\chi$ carries either a 
flavour index~\cite{Agrawal:2011ze, Kile:2013ola, Agrawal:2014una} 
or lepton number (leptophilic). In the former case the DM $\chi$'s interaction is restricted to only the
third generation of quarks or to top-quarks only. In the latter case the mediator carries leptonic number and the DM 
$\chi$ interacts only with the leptons~\cite{Agrawal:2014ufa, Freitas:2014jla}. 
In both these cases, the stringent bounds from the direct detection data are evaded.
Even if the DM interacts with only leptons the spin-independent $\chi$-nucleon interactions can still be generated and there
will be loop induced DM-hadron interaction where photon from virtual leptons can couple to the nucleus charge. This
contribution is however, subdominant being proportional to $\alpha_{\rm EM}^2$. It has recently been 
shown~\cite{Kilic:2015vka} in
direct detection searches at the LHC that a top flavoured DM of mass $> 200$~GeV can be consistent with current
bounds from direct detection experiments and relic density constraints.         
\end{itemize}



\acknowledgments
AG would like to thank the Mandelstam Institute for Theoretical Physics (University of the Witwatersrand) for their support 
and hospitality during his visit.	




\begin{thebibliography}{99}

\bibitem{Ade:2015xua} 
  P.~A.~R.~Ade {\it et al.} [Planck Collaboration],
  arXiv:1502.01589 [astro-ph.CO].
  
\bibitem{Goldberg:1983nd} 
  H.~Goldberg,
  Phys.\ Rev.\ Lett.\  {\bf 50}, 1419 (1983)
  Erratum: [Phys.\ Rev.\ Lett.\  {\bf 103}, 099905 (2009)].
  
\bibitem{Ellis:1983ew} 
  J.~R.~Ellis, J.~S.~Hagelin, D.~V.~Nanopoulos, K.~A.~Olive and M.~Srednicki,
  Nucl.\ Phys.\ B {\bf 238}, 453 (1984).
  
\bibitem{Kane:1993td} 
  G.~L.~Kane, C.~F.~Kolda, L.~Roszkowski and J.~D.~Wells,
  Phys.\ Rev.\ D {\bf 49}, 6173 (1994)
  [hep-ph/9312272].
\bibitem{Kolb:1983fm} 
  E.~W.~Kolb and R.~Slansky,
  Phys.\ Lett.\ B {\bf 135}, 378 (1984).
  
\bibitem{Cheng:2002ej} 
  H.~C.~Cheng, J.~L.~Feng and K.~T.~Matchev,
  Phys.\ Rev.\ Lett.\  {\bf 89}, 211301 (2002)
  [hep-ph/0207125].
  
\bibitem{Birkedal:2006fz} 
  A.~Birkedal, A.~Noble, M.~Perelstein and A.~Spray,
  Phys.\ Rev.\ D {\bf 74}, 035002 (2006)
  [hep-ph/0603077].
  
\bibitem{Beltran:2008xg} 
  M.~Beltran, D.~Hooper, E.~W.~Kolb and Z.~C.~Krusberg,
  Phys.\ Rev.\ D {\bf 80}, 043509 (2009)
  [arXiv:0808.3384 [hep-ph]].
  
\bibitem{Yu:2011by} 
  Z.~H.~Yu, J.~M.~Zheng, X.~J.~Bi, Z.~Li, D.~X.~Yao and H.~H.~Zhang,
  Nucl.\ Phys.\ B {\bf 860}, 115 (2012)
  [arXiv:1112.6052 [hep-ph]].
  
\bibitem{Goodman:2010yf} 
  J.~Goodman, M.~Ibe, A.~Rajaraman, W.~Shepherd, T.~M.~P.~Tait and H.~B.~Yu,
  Phys.\ Lett.\ B {\bf 695}, 185 (2011)
  [arXiv:1005.1286 [hep-ph]].
  
\bibitem{Goodman:2010ku} 
  J.~Goodman, M.~Ibe, A.~Rajaraman, W.~Shepherd, T.~M.~P.~Tait and H.~B.~Yu,
  Phys.\ Rev.\ D {\bf 82}, 116010 (2010)
  [arXiv:1008.1783 [hep-ph]].
  
\bibitem{Goodman:2010qn} 
  J.~Goodman, M.~Ibe, A.~Rajaraman, W.~Shepherd, T.~M.~P.~Tait and H.~B.~Yu,
  Nucl.\ Phys.\ B {\bf 844}, 55 (2011)
  [arXiv:1009.0008 [hep-ph]].
\bibitem{Fox:2011pm} 
  P.~J.~Fox, R.~Harnik, J.~Kopp and Y.~Tsai,
  Phys.\ Rev.\ D {\bf 85}, 056011 (2012)
  [arXiv:1109.4398 [hep-ph]].
  
\bibitem{Abdallah:2014hon} 
  J.~Abdallah {\it et al.},
  arXiv:1409.2893 [hep-ph].
  
\bibitem{Chang:2013oia} 
  S.~Chang, R.~Edezhath, J.~Hutchinson and M.~Luty,
  Phys.\ Rev.\ D {\bf 89}, no. 1, 015011 (2014)
  [arXiv:1307.8120 [hep-ph]].
\bibitem{DiFranzo:2013vra} 
  A.~DiFranzo, K.~I.~Nagao, A.~Rajaraman and T.~M.~P.~Tait,
  JHEP {\bf 1311}, 014 (2013)
  Erratum: [JHEP {\bf 1401}, 162 (2014)]
  [arXiv:1308.2679 [hep-ph]].
\bibitem{An:2013xka} 
  H.~An, L.~T.~Wang and H.~Zhang,
  Phys.\ Rev.\ D {\bf 89}, no. 11, 115014 (2014)
  [arXiv:1308.0592 [hep-ph]].
\bibitem{Papucci:2014iwa} 
  M.~Papucci, A.~Vichi and K.~M.~Zurek,
  JHEP {\bf 1411}, 024 (2014)
  [arXiv:1402.2285 [hep-ph]].

\bibitem{Baek:2015fma} 
  S.~Baek and Z.~F.~Kang,
  JHEP {\bf 1603}, 106 (2016)
  doi:10.1007/JHEP03(2016)106
  [arXiv:1510.00100 [hep-ph]].

\bibitem{Kile:2013ola} 
  J.~Kile,
  Mod.\ Phys.\ Lett.\ A {\bf 28}, 1330031 (2013)
  [arXiv:1308.0584 [hep-ph]].
  
\bibitem{Agrawal:2014una} 
  P.~Agrawal, B.~Batell, D.~Hooper and T.~Lin,
  Phys.\ Rev.\ D {\bf 90}, no. 6, 063512 (2014)
  [arXiv:1404.1373 [hep-ph]].

\bibitem{Englert:2016joy} 
  C.~Englert, M.~McCullough and M.~Spannowsky,
  Phys.\ Dark Univ.\  {\bf 14}, 48 (2016)
  doi:10.1016/j.dark.2016.09.002
  [arXiv:1604.07975 [hep-ph]].
    
\bibitem{Belanger:2014vza} 
  G.~Bélanger, F.~Boudjema, A.~Pukhov and A.~Semenov,
  Comput.\ Phys.\ Commun.\  {\bf 192}, 322 (2015)
  [arXiv:1407.6129 [hep-ph]].

\bibitem{Alloul:2013bka}
  A.~Alloul, N.~D.~Christensen, C.~Degrande, C.~Duhr and B.~Fuks,
  Comput.\ Phys.\ Commun.\  {\bf 185}, 2250 (2014)
  [arXiv:1310.1921 [hep-ph]].
  
  
\bibitem{Akerib:2013tjd} 
  D.~S.~Akerib {\it et al.} [LUX Collaboration],
  Phys.\ Rev.\ Lett.\  {\bf 112}, 091303 (2014)
  [arXiv:1310.8214 [astro-ph.CO]].
  

\bibitem{Aprile:2013doa} 
  E.~Aprile {\it et al.} [XENON100 Collaboration],
  Phys.\ Rev.\ Lett.\  {\bf 111}, no. 2, 021301 (2013)
  [arXiv:1301.6620 [astro-ph.CO]].
  
\bibitem{Ackermann:2015zua} 
  M.~Ackermann {\it et al.} [Fermi-LAT Collaboration],
  Phys.\ Rev.\ Lett.\  {\bf 115}, no. 23, 231301 (2015)
  [arXiv:1503.02641 [astro-ph.HE]].

\bibitem{Haisch:2016usn} 
  U.~Haisch, F.~Kahlhoefer and T.~M.~P.~Tait,
  Phys.\ Lett.\ B {\bf 760}, 207 (2016)
  doi:10.1016/j.physletb.2016.06.063
  [arXiv:1603.01267 [hep-ph]].
  
\bibitem{Askew:2014kqa} 
  A.~Askew, S.~Chauhan, B.~Penning, W.~Shepherd and M.~Tripathi,
  Int.\ J.\ Mod.\ Phys.\ A {\bf 29}, 1430041 (2014)
  [arXiv:1406.5662 [hep-ph]].
  
\bibitem{Khachatryan:2014rra} 
  V.~Khachatryan {\it et al.} [CMS Collaboration],
  Eur.\ Phys.\ J.\ C {\bf 75}, no. 5, 235 (2015)
  [arXiv:1408.3583 [hep-ex]].
  
\bibitem{CMS:2013gea} 
  CMS Collaboration [CMS Collaboration],
  CMS-PAS-SUS-13-012.
  
\bibitem{TheATLAScollaboration:2013fha} 
  The ATLAS collaboration [ATLAS Collaboration],
  ATLAS-CONF-2013-047.

\bibitem{Alwall:2014hca}
  J.~Alwall {\it et al.},
  JHEP {\bf 1407}, 079 (2014)
  [arXiv:1405.0301 [hep-ph]].

\bibitem{Drees:1993bu} 
  M.~Drees and M.~Nojiri,
  Phys.\ Rev.\ D {\bf 48}, 3483 (1993)
  [hep-ph/9307208].

\bibitem{Matsumoto:2016hbs} 
  S.~Matsumoto, S.~Mukhopadhyay and Y.~L.~S.~Tsai,
  arXiv:1604.02230 [hep-ph].
    
\bibitem{Agrawal:2011ze} 
  P.~Agrawal, S.~Blanchet, Z.~Chacko and C.~Kilic,
  Phys.\ Rev.\ D {\bf 86}, 055002 (2012)
  [arXiv:1109.3516 [hep-ph]].
  
\bibitem{Agrawal:2014ufa} 
  P.~Agrawal, Z.~Chacko and C.~B.~Verhaaren,
  JHEP {\bf 1408}, 147 (2014)
  [arXiv:1402.7369 [hep-ph]].
  
\bibitem{Freitas:2014jla} 
  A.~Freitas and S.~Westhoff,
  JHEP {\bf 1410}, 116 (2014)
  [arXiv:1408.1959 [hep-ph]].
  
\bibitem{Kilic:2015vka} 
  C.~Kilic, M.~D.~Klimek and J.~H.~Yu,
  Phys.\ Rev.\ D {\bf 91}, no. 5, 054036 (2015)
  [arXiv:1501.02202 [hep-ph]].


  



\end{thebibliography}
\end{document}